\documentclass[aip,pop,preprint,superscriptaddress,showpacs,10pt]{revtex4-1}
\usepackage{amsfonts}
\usepackage{amsmath}
\usepackage{amssymb}
\usepackage{graphicx}
\usepackage{helvet}
\usepackage{hyperref}
\usepackage{color}

\begin{document}

\title{Modulational instability and nonlinear evolution of two-dimensional
electrostatic wave packets in ultra-relativistic degenerate dense
plasmas}

\author{Amar Prasad Misra}

\email{apmisra@visva-bharati.ac.in; apmisra@gmail.com}

\altaffiliation{On leave from Department of Mathematics, Siksha Bhavana, Visva-Bharati University, Santiniketan-731 235, India.}

\affiliation{Department of Physics, Ume\aa\ University, SE-901 87 Ume\aa ,
Sweden}

\author{Padma Kant Shukla}

\email{ps@tp4.rub.de; profshukla@yahoo.com}

\affiliation{Department of Physics, Ume\aa\ University, SE-901 87 Ume\aa ,
Sweden}

\affiliation{RUB International Chair, International Centre for Advanced Studies
in Physical Sciences, Faculty of Physics \& Astronomy, Ruhr University
Bochum, D-44780 Bochum, Germany}

\received{06 December 2010}
\revised{17 March 2011}
\begin{abstract}
We consider the nonlinear propagation of electrostatic wave packets
in an ultra-relativistic (UR) degenerate  dense electron-ion plasma, whose
dynamics is governed by the nonlocal two-dimensional nonlinear
Schr{\"o}dinger-like equations. The coupled set of equations are then used
to study the modulational instability (MI) of a uniform wave train
to an infinitesimal perturbation of multi-dimensional form. The condition
for the MI is obtained, and it is shown that the nondimensional parameter, $\beta\propto\lambda_C n_0^{1/3}$ (where 
$\lambda_C$ is the reduced Compton wavelength and $n_0$ is the particle number density), associated with the UR pressure of     degenerate electrons, shifts the stable (unstable) regions at $n_{0}\sim10^{30}$ cm$^{-3}$ to unstable (stable) ones at higher densities, i.e.  $n_{0}\gtrsim7\times10^{33}$.  It is also found that {the} higher the values of  $n_{0}$,
the lower is the growth rate of MI with cut-offs at lower wave numbers
of modulation. Furthermore, the dynamical evolution of the wave packets is 
studied numerically. We show that  either they disperse away or they blowup in
a finite time, when the wave action is below or above the threshold.
The results could be useful for understanding the properties of modulated wave packets and their multi-dimensional evolution in   UR degenerate dense plasmas, such as those in the interior of white dwarfs {and/or} pre-Supernova stars. 
\end{abstract}



\maketitle


\section{Introduction}

As is known, compact astrophysical objects, e.g. massive white dwarfs
{and/or}  the core of pre-Supernova stars, are supported by the pressure
of degenerate electrons, i.e., in their interiors the particle number
density is extremely high \cite{compact-objects,compact-objects-Pressure}. 
So, the Fermi energy can be much larger than
the thermal energy (e.g. typical electron Fermi energy, $E_{F}\sim1$
MeV corresponds to a Fermi temperature $T_{F}\sim10^{10}$ K). Thus,
the thermal pressure of electrons may be negligible compared to the
Fermi pressure. Such degenerate electrons may, however, be either
nonrelativistic, somewhat relativistic or ultra-relativistic. In the
latter case, the speed of electrons can approach the speed of light
in vacuum $(c)$, and  the equation of state for electrons can then be
written as \cite{compact-objects-Pressure,Ultra-relativistic-photon,Ultra-relativistic-pressure} $P_{e}=\left(3/\pi\right)^{1/3}\left(\hbar c/8\right)n_{e}^{4/3}$
whenever $E_{F}\gg m_{e}c^{2}$. Here, $\hbar$ is the reduced Planck's
constant, $n_{e}$ is the electron number density and $m_{e}$ is
the electron mass. Such ultra-relativistic degenerate (URD) electrons
are, indeed, ubiquitous in many other astrophysical environments including
neutron stars {and} magnetars \cite{compact-objects,compact-objects-Pressure,Quantum-plasma-review}. 

Now, in dealing with those compact objects like white dwarfs, their interiors
may be considered as a plasma system (e.g., carbon-oxygen white
dwarf in a thermonuclear Supernova explosion) consisting of positively
charged ions (nuclei) providing almost all the mass (inertia) and
none of the pressure, as well as electrons providing all the pressure (restoring
force), but none of the mass (inertialess). So, interiors of such
compact objects  provide us a cosmic laboratory for studying the
properties of such plasmas  as well as nonlinear collective oscillations
under extreme conditions, i.e. at the relativistically degenerate
dense states with higher densities \cite{Quantum-plasma-review} $(10^{6}-10^{9}$g/cm$^{3})$. Recent investigations
along these lines indicate that such URD dense plasmas can support
the propagation of solitary waves as well as double layers at different
length scales of excitation \cite{Ultra-relativistic-solitary}.

On the other hand, the nonlinear propagation of wave packets in plasmas
is generically subject to their amplitude modulation due to the carrier
wave self-interaction, i.e., a slow variation of the wave envelope
due to nonlinearities (see, e.g. \cite{Misra1,Misra2,Misra3,Misra4}). Under certain conditions, the system's evolution
shows a modulational instability (MI), leading to the formation of
envelope solitons through the localization of wave energy. Such solitons, governed
by a (1+1)-dimensional nonlinear Schr{\"o}dinger equation (NLSE), are the
result of a balance between the nonlinearity (self-focusing) and the wave group dispersion.
When such a balance can be maintained dynamically, the solitons may exist
even under strong perturbations. On the contrary, the multi-dimensional [(2+1) or (3+1)-dimensional]
NLSE with cubic  and/ or quadratic (nonlocal) nonlinearities may no longer be integrable. In this case, the self-focusing effect can dominate over the dispersion (beyond a threshold intensity) leading to the formation of wave collapse or blowup of the wave amplitude \cite{collapse-review,collapse-plasmas}. 

The important issue for the {multi}-dimensional NLSE  may be that for a wide range of initial conditions, the system often exhibits collapse  in which a singularity of the wave field is formed in a finite time, instead of a stable mode propagation.  A collapsing wave packet thus self-focuses  in shorter scales and with higher amplitudes until other physical effects intervene to arrest it.  Such collapse phenomenon plays an
important role as an effective mechanism for the energy localization  in various branches of physics \cite{collapse-review,collapse-plasmas}. {In this context, it could be}  a central problem of finding a proper initial condition which leads to wave collapse. This is, one of our goals of the present work in the description of Davey-Stewartson II (DS II)-like equations  \cite{DS-original} for the wave packets. Several authors have investigated such collapse theory, e.g. in {the} context of hydrodynamics \cite{collapse-hydrodynamics}, in nonlinear optics \cite{collapse-optics},  in Bose-Einstein condensates \cite{collapse-bose-einstein} as well as in plasma physics \cite{collapse-plasmas}.  However, to our knowledge, there are few theoretical studies for the MI (see, e.g. \cite{DS-plasma-cylindrical,DS-plasma-MI,DS-II-plasma-dromion}) and multi-dimensional evolution of electrostatic wave packets (see, e.g. \cite{DS-II-plasma-dromion}) in plasmas, which are described by the DS II-like equations. The latter generalize the (1+1)-dimensional NLSE with a nonlocal (quadratic) nonlinear term associated with the static field in the plasma. 

 Furthermore, since quadratic nonlinearities are known to be collapse-free, multi-dimensional NLSE, where the cubic and quadratic (nonlocal) nonlinearities compete, can also support coherent structures, i.e. dromion-like solutions ({unlike} solitons the dromions can have inelastic collisions
and can transfer mass or energy) which may  either decay due to the dispersion to be enhanced by the static field or exhibit blowup due to nonlinearity, in a finite time \cite{DS-II-numerical1,DS-II-numerical2}.  However, their applications in physical systems, especially in plasmas are not yet fully understood or less developed till now. 

In this article, our purpose is to consider the propagation of two-dimensional (2D) electrostatic wave packets (EWPs) in a URD dense plasma. We provide a
general criterion for the MI  of a plane wave packet as well as the instability growth rate. We show that as one
approaches the higher density regimes, the stability of the wave increases with lower growth rates at lower wave numbers of modulation. Furthermore,
the 2D evolution of the nonlocal NLSEs exhibit dromion-like solutions which either decay by the wave dispersion or blowup due
to wave nonlinearity in a finite interval of time. The latter are in qualitative agreement with the results \cite{DS-II-numerical1,DS-II-numerical2} already found in DS II equations for water waves with finite depth \cite{DS-original}.

\section{Basic equations and derivation of the evolution equations}

Let us consider the propagation of EWPs in a 2D dense plasma composed
of inertialess URD ultra-cold electrons
and inertial ultra-cold ions. Any speed involved in the plasma flow
is assumed to be much lower than the ion-acoustic speed. At equilibrium,
{both  species} have equal number density, say $n_{0}$. Assuming
further that the plasma is collisionless and unmagnetized, the basic
normalized equations then read \cite{Ultra-relativistic-solitary}
\begin{eqnarray}
 &  & \frac{\partial n_{i}}{\partial t}+\nabla\cdot(n_{i}\mathbf{v})=0,\label{e1}\\
 &  & \frac{\partial\mathbf{v}}{\partial t}+(\mathbf{v}\cdot\nabla)\mathbf{v}=-\nabla\phi,\label{e2}\\
 &  & 0=\nabla\phi-\frac{3\beta}{4n_{e}}\nabla n_{e}^{4/3},\label{e3}\\
 &  & \nabla^{2}\phi=n_{e}-n_{i},\label{e4}\end{eqnarray}
 where $\nabla\equiv(\partial/\partial x,\partial/\partial y),$ $n_{i}$
is the ion number density normalized by $n_{0}$, {$\mathbf{v}\equiv(v_{x},v_{y})$}
is the ion velocity normalized by $C_{i}=\sqrt{Z_{i}m_{e}c^{2}/m_{i}}$
with $Z_{i}$ denoting the ion charge state, $m_{i}$ {is} the ion
 mass and $c$ {is} the speed of light in vacuum. Also, $\phi$ is
the electrostatic wave potential normalized by $Z_{i}m_{e}c^{2}/e$
with $e$ denoting the elementary charge. The space and time variables
are respectively normalized by the screening length $\lambda_{s}=\sqrt{m_{e}c^{2}/4\pi n_{0}e^{2}}$
and the ion plasma period, $\omega_{pi}^{-1}=\sqrt{m_{i}/4\pi n_{0}Z_{i}^{2}e^{2}}$.
Moreover, in Eq. (\ref{e3}), $\beta=\lambda_{C}\sqrt[3]{n_{0}/72\pi}$  where
$\lambda_{C}\equiv\hbar/mc$ is the reduced Compton wavelength, and
we have used the same equation of state {for} $P_{e}$ as described in the previous section
relevant for URD electrons in dense plasmas.
Furthermore, the nondimensional parameter, {$\beta\gtrless1$ for $n_{0}\gtrless n_{c}\thickapprox3.94\times10^{33}$cm$^{-3}$ and $\beta=1$ for $n_{0}=n_{c}$}, i.e. higher values of $\beta$ represent the higher-density regimes. 

Next, we consider the time evolution of a wave packet of electrostatic
perturbations which occur along the $x$-axis. The initial wave packets
of perturbation are basically modulated by the nonlinear carrier wave
self-interactions. Then if one observes the wave packets from a coordinate
frame moving with a group speed $v_{g}$, to be obtained from the
linear dispersion relation (as given below) of Eqs. (\ref{e1})-(\ref{e4}),
then the time variation of the wave packets {looks} slow, and so
the space and the time variables can be stretched as\begin{equation}
\xi=\epsilon(x-v_{g}t),\eta=\epsilon y,\tau=\epsilon^{2}t,\label{e6}\end{equation}
 where $\epsilon$ is a small parameter representing the strength
of the wave amplitude. {We are interested} in the modulation of
a plane wave as the carrier wave with wave {number} and frequency $k$ and $\omega$ respectively, then  following the standard reductive perturbation technique
(RPT) (see, e.g. \cite{DS-II-plasma-dromion}), the dynamical variables can be expanded as\begin{eqnarray}
 &  & n_{e,i}=1+\sum_{n=1}^{\infty}\epsilon^{n}\sum_{l=-\infty}^{\infty}n_{(e,i)l}^{(n)}(\xi,\eta,\tau)\exp\left[i(kx-\omega t)l\right],\label{e7}\\
 &  & v_{x,y}=\sum_{n=1}^{\infty}\epsilon^{n}\sum_{l=-\infty}^{\infty}v_{(x,y)l}^{(n)}(\xi,\eta,\tau)\exp\left[i(kx-\omega t)l\right],\label{eq:e8-1}\\
 &  & \phi=\sum_{n=1}^{\infty}\epsilon^{n}\sum_{l=-\infty}^{\infty}\phi_{l}^{(n)}(\xi,\eta,\tau)\exp\left[i(kx-\omega t)l\right],\label{e9}\end{eqnarray}
 where $n_{(e,i)l}^{(n)},$ $v_{(x,y)l}^{(n)}$ and $\phi_{l}^{(n)}$
should satisfy $S_{-l}^{(n)}=S_{l}^{(n)\ast}$ because of the {reality}
condition for the physical variables. Here, the asterisk denotes the
complex conjugate of the corresponding quantity. Note that the
group speed, $v_{g}$ is now normalized by $C_{i}$ (smaller than the ion-sound speed),  the wave
frequency $\omega$ and the wave number  $k$ {are normalized by}  $\omega_{pi}$
and $\lambda_{s}$ respectively.

We now substitute the {expressions (\ref{e7})-(\ref{e9})}
into the basic Eqs. (\ref{e1})-(\ref{e4}) and equate the terms in different
powers of $\epsilon$. We shall, however, omit the detail calculations,
since the procedure is quite standard and follows the usual RPT. In the lowest order of $\epsilon$, we obtain for $n=1$, $l=1$ the
linear dispersion relation [since we are considering the modulation
of a plane wave, $n_{(e,i)l}^{(1)},$ $v_{(x,y)l}^{(1)}$ and $\phi_{l}^{(1)}$
are all set to zero except {for $l=\pm1$]}
 \begin{equation}
\omega^{2}=\frac{\beta k^{2}}{1+\beta k^{2}}.\label{e10}
\end{equation}
 We find that the dispersion equation (\ref{e10}) has the similar form with {that derived by Kako \textit{et al} \cite{classical-2D-MI}} for classical
plasmas comprising cold ions and isothermal electrons whenever one replaces
 $\beta$ by unity. However, one should note that
 the {case  $\beta=1$}, which represents the higher density regimes where $n_{0}=n_{c}$, 
is not applicable for classical isothermal plasmas. Equation (\ref{e10})
shows that the wave always propagates with a frequency below the ion-plasma
frequency regardless of the values of $\beta$ and $k$. In the short-wavelength
limit, the frequency of the wave {approaches  unity} (the upper limit
of the normalized $\omega$), whereas in the {long-wavelength limit}, it approaches
 a zero value. Furthermore, the limit $\beta k^{2}\ll1$ is not
admissible because otherwise, $\omega$ will  be small to provide weak dispersion, and
 the soliton formation (Korteweg-de Vries soliton) of
the carrier wave will be a lower-order process than the MI of the
envelope to be studied here. Moreover, since $k$ is normalized by the
inverse of the screening length $\lambda_{s}$, the values of $k>2\pi$
{are} also inadmissible, otherwise the wavelength would become smaller
than the screening length. As a result, the plasma collective behaviors
might disappear.

Now, proceeding in the same way as of Refs. \cite{DS-plasma-cylindrical,DS-plasma-MI,DS-II-plasma-dromion}, i.e., considering the second
harmonic modes obtained in terms of $\phi_{1}^{(2)}$, $\partial\phi_{1}^{(1)}/\partial\xi$, $\partial\phi_{1}^{(1)}/\partial\eta$
for $n=2$, $l=1$ and $\left[\phi_{1}^{(1)}\right]^{2}$ for $n=2,l=2$
as well as the zeroth-harmonic modes {appearing} due to the nonlinear
self-interaction of the modulated carrier waves, and finally considering
the equations for $n=3$, $l=1$, we obtain the following 2D nonlocal NLSEs
 for the propagation of modulated wave packets in URD  {dense plasmas} 
\begin{eqnarray}
 &  & i\frac{\partial\phi}{\partial\tau}+P_{1}\frac{\partial^{2}\phi}{\partial\xi^{2}}+P_{2}\frac{\partial^{2}\phi}{\partial\eta^{2}}+Q_{1}|\phi|^{2}\phi+Q_{2}\psi\phi=0\label{eq:g1},\\
 &  & R\frac{\partial^{2}\psi}{\partial\xi^{2}}+\frac{\partial^{2}\psi}{\partial\eta^{2}}=S\frac{\partial^{2}|\phi|^{2}}{\partial\xi^{2}}\label{eq:g2},\end{eqnarray}
 where $\phi\equiv\phi_{1}^{(1)}$, $\psi\equiv\int\partial_{\xi}v_{y0}^{(2)}\partial\eta$
and the coefficients written only in terms of $k$ and the parameter $\beta$ are given by

\begin{eqnarray}
& & P_{1}=-\frac{3k\beta^{3/2}}{2\left(1+\beta k^{2}\right)^{5/2}},\hskip 10pt P_{2}=\frac{\sqrt{\beta}}{2k\left(1+\beta k^{2}\right)^{3/2}},\label{eq:c1}\\
& & Q_{1}=\frac{k\sqrt{1+\beta k^{2}}}{9\sqrt{\beta}}\left[-\frac{3}{2}+\frac{7\beta^{4}}{3\left(1+\beta k^{2}\right)^{2}}\right]\left(2+3k^{2}+\frac{2}{\beta^{2}k^{2}}\right)\notag\\
& & \hskip20pt+\frac{k\left(1+\beta k^{2}\right)^{3/2}}{\beta^{3/2}}\left[1+\frac{\beta^{3}\left(5-2/\beta^{3}\right)}{3\left(1+\beta k^{2}\right)^{3}}\right]
-\frac{k\left(-2+3\beta+3\beta^{2}k^{2}\right)}{9\beta^{5/2}\left(1+\beta k^{2}\right)^{3/2}},\label{eq:c2}\\
&& Q_{2}=\frac{k\left(5-3\beta k^{2}\right)}{6\sqrt{\beta}\left(1+\beta k^{2}\right)^{3/2}},\hskip10pt R=1-\frac{1}{\left(1+\beta k^{2}\right)^{3}},\label{eq:c3}\\
&& S=-\frac{1+\beta k^{2}}{\beta^{3/2}}\left[3+\beta k^{2}-\frac{2}{3\left(1+\beta k^{2}\right)^{3}}\right].\label{eq:c4}
\end{eqnarray}
 We note that since, $P_{1}<0$, $P_{2}>0$ and $R>0$, the  NLSEs (\ref{eq:g1})
and (\ref{eq:g2}) are in the form of second type (hyperbolic-elliptic)
Davey-Stewartson, i.e. DS II-like equations. Furthermore,  { $S$ is always negative,  $Q_{2}\gtrless0$
for $\beta k^{2}\lessgtr5/3$, $Q_2=0$ for $\beta k^{2}=5/3$} and $Q_{1}$ can be either positive
or negative depending on the values of both  $k$ and $\beta$ . In the  {case  $Q_{2}=0$}
 for which Eqs. (\ref{eq:g1}) and (\ref{eq:g2})
are reduced to 2D NLSE, the modulated wave can be shown to
be stable. This will be clear from the next section dealing with the MI. We will see that for  values of { $\beta>1$} (which corresponds to higher density regimes) and  small wave numbers, i.e. $k<1$ fulfilling $\beta k^2\lessapprox3$, the wave tends to become stable. 

Thus, we have obtained a new set of nonlocal NLSEs, which describe the slow modulation of EWPs in 2D URD dense plasmas. The coefficients $P_{1}$, $P_{2}$
appear due to the wave group dispersion and the 2D evolution of the EWPs. One of the  nonlinear coefficients, $Q_{1}$ (cubic) is due to the carrier
wave self-interaction originating from the zeroth harmonic modes (or
slow modes)  and the other nonlocal coefficient (quadratic) $Q_{2}$ comes from the coupling between the dynamical
field associated with the first harmonic (with a `cascaded' effect from the second harmonic) and a static field generated due to the
mean motion (zeroth harmonic) in  plasmas.

\section{modulational instability and  growth rate}

In this section, we consider the modulation of a plane wave solution of Eqs. (\ref{eq:g1})
and (\ref{eq:g2}) for $\phi$ with a constant amplitude $\phi_{0}.$
The boundary conditions, namely $\phi,\psi\rightarrow0$ as $\xi,\eta\rightarrow\infty$
used before must now be relaxed, since the  wave packet is still  not modulated
and the solution is not unique. Note that the choice of $\psi=\psi_{0}$ is immaterial as the
stability criterion does not depend on it.  Thus, we can represent the plane wave solution
as $\phi=\phi_{0} \exp\left[i(k_{\xi}\xi+k_{\eta}\eta-\Delta\tau)\right]$,
and $\psi=\psi_{0}$  with $\Delta=P_{1}k_{\xi}^{2}+P_{2}k_{\eta}^{2}-Q_{1}\phi_{0}^{2}-Q_{2}\psi_{0}$, 
where $k_{\xi}$, $k_{\eta}$, $\psi_{0}$ are all real constants. Next, to study
the stability of this solution we modulate the amplitude  as $\phi=\left(\phi_{0}+\phi_{m}\right)\exp\left[i(k_{\xi}\xi+k_{\eta}\eta-\Delta\tau)\right]$, 
$\psi=\psi_{0}+\psi_{m}$, where $\phi_{m},\psi_{m}\propto\mbox{Re}\left[\exp i(K_{1}\xi+K_{2}\eta-\Omega\tau)\right]$
and $K,$ $\Omega$ are respectively the wave number and the wave
frequency of modulation. Looking for the nonzero solution of the small
amplitude  perturbations, we  obtain from Eqs.
(\ref{eq:g1}) and (\ref{eq:g2}) the following dispersion relation for the modulated  {wave packet}
 \begin{equation}
\Omega^{2}=\left(P_{1}K_{1}^{2}+P_{2}K_{2}^{2}\right)^{2}\left(1-{K_{c}^{2}}/{K^{2}}\right),\label{e18}\end{equation}
 where \begin{equation}
K_{c}^{2}=\frac{2\phi_{0}^{2}\left[\left(R+\zeta^{2}\right)Q_{1}+SQ_{2}\right]\left(1+\zeta^{2}\right)}{\left(P_{1}+P_{2}\zeta^{2}\right)\left(R+\zeta^{2}\right)},\label{eq:Kc}\end{equation}
 in which  $\zeta\equiv\tan\theta=K_{2}/K_{1}$. Equation (\ref{e18}) shows that the MI
sets in for a wave number satisfying $K<K_{c}$ or, for all wavelengths above
the threshold, $\lambda_{c}=2\pi/K_{c}$, provided the right-hand
side of Eq. (\ref{eq:Kc}) is positive, i.e. 
\begin{equation}
\frac{\left(R+\zeta^{2}\right)Q_{1}+SQ_{2}}{\left(P_{1}+P_{2}\zeta^{2}\right)\left(R+\zeta^{2}\right)}>0.\label{eq:cond1}
\end{equation}
 Since $R>0$, the condition (\ref{eq:cond1}) reduces to

\begin{equation}
\Lambda\equiv\left[\left(R+\zeta^{2}\right)Q_{1}+SQ_{2}\right]\left(P_{1}+P_{2}\zeta^{2}\right)>0.\label{eq:cond2}\end{equation}
The wave packet is otherwise (i.e., for $K>K_c$) said to be  stable under the modulation. The instability growth rate (letting
$\Omega=i\Gamma)$ can be obtained from Eq. (\ref{e18}) as
 \begin{equation}
\Gamma=\frac{K^{2}\left(P_{1}+P_{2}\zeta^{2}\right)}{1+\zeta^{2}}\sqrt{\frac{K_{c}^{2}}{K^{2}}-1}.\label{eq:growth}
\end{equation}
Clearly, the maximum growth rate, achieved at $K=K_{c}/\sqrt{2}$,  is $\Gamma_{\text{max}}=\left[\left(R+\zeta^{2}\right)Q_{1}+SQ_{2}\right]|\phi_{0}|^{2}/\left(R+\zeta^{2}\right)$. 
 
Next, we numerically investigate the condition of MI given by Eqs. (\ref{eq:cond1}) or (\ref{eq:cond2}) as well as the instability growth rate, $\Gamma$
given above. The condition of MI not only depends on the carrier wave
number $k$ and the density dependent parameter $\beta$ arising due to the ultra-relativistic pressure of degenerate electrons, but also
on the obliqueness parameter $\theta$ of the modulational wave number $K$
with the $\xi$-axis due to 2D perturbation. The stable $\left(\Lambda<0\right)$
and unstable $\left(\Lambda>0\right)$ regions are thus shown in the
$k\theta$ plane in Fig. 1 for different values of $\beta$ that correspond to different density regimes. We find
that the stable and unstable regions are completely divided into two parts at comparatively lower as well as higher density plasmas, i.e. at
 $n_{0}\sim10^{30}$ cm$^{-3}$ {[}see Fig. 1(a){]}
and at $n_{0}=5\times10^{33}, 7\times10^{33}$ and $10^{34}$ cm$^{-3}$
{[}see Figs. 1(d)-1(f){]}. 
Separation of such  regions are also observed at the intermediate  densities [Figs. 1(b) and (c)]. As seen from these figures, there are basically four regions (two for each) for the stable and unstable waves, and for a wide range of values of $k$ and $\tan\theta$.
 From  Figs. 1(a) and 1(e) or 1(f), one observes that some part of the regions in which the wave was stable (unstable) at lower density 
now shifts to  unstable (stable) region at higher values of the same. Obviously, the parameter responsible for this shift is $\beta$  due to the consideration of ultra-relativistic degenerate electrons.

Again, { $P_{2}$, $R$ are always positive and $Q_{2}>0$ for $\beta k^{2}<5/3$}.
Also, $S<0$ and $P_{2}>\mid P_{1}\mid$ for $\beta k^{2}>1/2$. Then, in the regime satisfying 
$1/2<\beta k^{2}<5/3$,  the instability condition (\ref{eq:cond1}) or (\ref{eq:cond2}) depends mainly on the sign and magnitude of the coefficient $Q_{1}$. Physically, this implies that the EWPs become stable or unstable in the said regime due to the nonlinear self-interactions originating from the second harmonic modes as well as from the zeroth
harmonic modes (or slow modes). In particular,  under the horizontal modulation $\left(K_{1}=0\right)$
the instability condition (\ref{eq:cond1}) reduces to $\Lambda=Q_{1}/P_{2}$ which can be either positive
or negative depending on the sign of $Q_{1}$ (since $P_2>0$). We also find that for a particle
density to vary in $10^{32}-10^{33}$ cm$^{-3}$ and for carrier 
wave numbers $k<1$, $Q_{1}$ is negative, and then the modulated wave is said to be  
stable. On the other hand, under the longitudinal modulation, i.e.
$K_{2}=0$, we have $\Lambda=\left(Q_{1}R+SQ_{2}\right)/P_{1},$ which
may, however, be positive almost everywhere in the above regime, and hence the instability.
  These results are in qualitative agreement with
the classical ones considered before by Nishinari et al \cite{DS-II-plasma-dromion}
in the description of DS II equations for EWPs. However, the general situation
is quite different as clear from Fig. 1. 

Now, the MI growth rate can be calculated from Eq. (\ref{eq:growth})
for a fixed value of the carrier wave number. Figure 2 shows that
higher the density regimes, the lower is the growth rate of instability
with cut-offs at lower wave numbers of modulation. It seems that the parameter $\beta$ plays almost the similar role of dispersion 
 as  the quantum diffraction associated with the Bohm potential plays for modulated wave packets in quantum plasmas \cite{Misra1,Misra2,Misra3,Misra4}. This is expected as we have seen from Fig. 1 that the waves tend to become stable at
higher  number densities.  {Thus,  from this section we conclude that the number densities of ultra-relativistic ultra-cold  electrons   significantly modify the stability and instability regions in the $k\theta$-plane  as well as reduce the instability growth rate for the wave packets}. In the next section, we will  numerically investigate the dynamical evolution of the EWPs that undergo MI in the regimes discussed above. We will see that the situation is quite distinctive from the one-dimensional case in which an exact balance between the self-focusing and the dispersion can happen  to form envelope solitons on stable wave propagation.   

%
\begin{figure}[btp]
\includegraphics[width=7in,height=4in,trim=0.0in 4in 0in 4in]{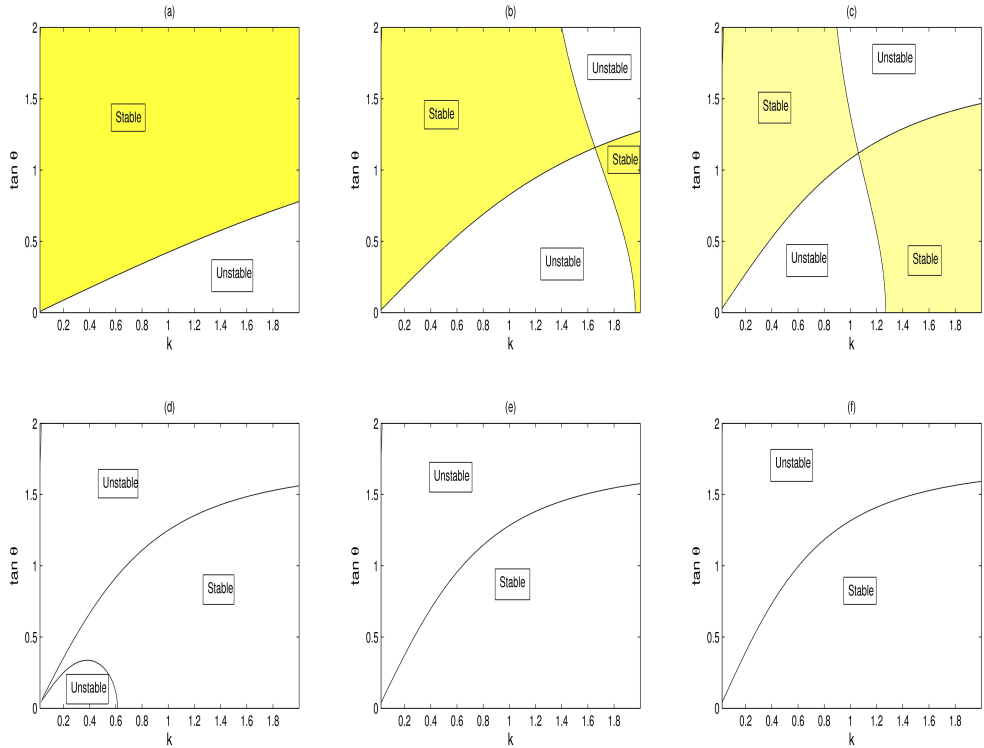} 
\caption{ (Color online) Contour plots of $\Lambda=$const. against $k$ and $\tan \theta$. The stable (shaded and/or wherever mentioned) and unstable 
(white and/or wherever mentioned) regions are 
shown in the $k\theta$-plane. The panels (a) to (f) represent respectively
the regions corresponding to the densities, $n_{0}$ (cm$^{-3}$)$=10^{30}$,
$10^{32}$, $10^{33}$,  $5\times10^{33}$, $7\times10^{33}$ and $10^{34}$.
The  critical value at which the stable (unstable)
regions are shifted to unstable (stable) ones is $n_0=7\times10^{33}$ cm$^{-3}$, and above which the stability region increases slightly with $n_0$ [see Fig. 1(f)]. The panels (a), (e) and (f) show
that the regions of stability and instability
are divided  into two parts at both lower and higher densities. }
\end{figure}
\begin{figure}[btp]
\includegraphics[width=5in,height=3in,trim=0.0in 3.5in 0in 4in]{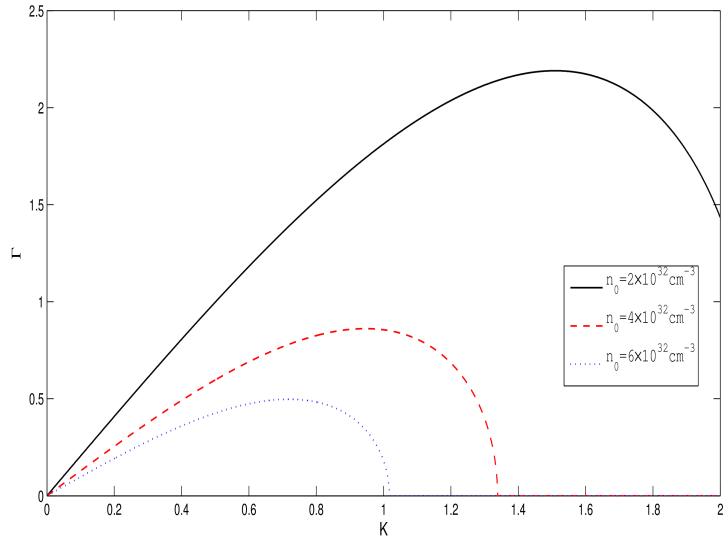} 
\caption{ (Color online) The modulational instability growth rate, $\Gamma$ is shown with respect to the wave number of modulation, $K$ for a fixed $k=0.6$, 
$\zeta=0.2$ and $\phi_0=0.1$. This shows that  {the} higher the particle density, the lower is the growth rate, $\Gamma$ with cut-offs at lower $K$.}
\end{figure}
\section{2D evolution of the nonlocal equations}

In the evolution of EWPs described by the Eqs. (\ref{eq:g1}) and (\ref{eq:g2}) we first obtain some analytic conditions for the wave collapse to occur within a finite time. Note that since the evolution
equations are of the DS II-type, we can not derive the criteria that
are sufficient to ensure collapse by the Virial theorem or else \cite{collapse-review}. However,
we will present some conditions according to Berkshire and Gibbon \cite{collapse-criteria}.
To that end, we first see that the integrals of motion are the wave
action $N$ and the Hamiltonian $\mathcal{H}$ where \begin{equation}
N=\iint\left\vert \phi\right\vert ^{2}d\xi d\eta,\label{eq:wave-action}\end{equation}

\begin{align}
\mathcal{H} & =\iint\left[P_{1}\left\vert \frac{\partial\phi}{\partial\xi}\right\vert ^{2}+P_{2}\left\vert \frac{\partial\phi}{\partial\eta}\right\vert ^{2}+\frac{1}{2}Q_{1}\left\vert \phi\right\vert ^{4}\right]d\xi d\eta\notag\\
 & -\frac{Q_{2}}{2S}\iint\left[R\left(\frac{\partial^{2}u}{\partial\xi^{2}}\right)^{2}+\left(\frac{\partial^{2}u}{\partial\xi\partial\eta}\right)^{2}-2S\frac{\partial^{2}u}{\partial\xi^{2}}\left\vert \phi\right\vert ^{2}\right]d\xi d\eta,\label{Hamiltonian}\end{align}
 where $\partial^{2}u/\partial\xi^{2}\equiv\psi$. If $I$ is the
moment of inertia of a localized wave form, then by the Virial theorem
we have \cite{collapse-review}
 \begin{equation}
\frac{d^{2}I}{d\tau^{2}}\equiv\frac{d^{2}}{d\tau^{2}}\left[\iint\left(\frac{\xi^{2}}{P_{1}}+\frac{\eta^{2}}{P_{2}}\right)\left\vert \phi\right\vert ^{2}d\xi d\eta\right]=8\mathcal{H},\label{eq:virial}
\end{equation}
 which upon integration gives 
 \begin{equation}
\frac{dI}{d\tau}=8\mathcal{H}\tau+C_{1},
\end{equation}
 or, \begin{equation}
\left<I\right>\equiv\frac{I}{N}=\frac{4\mathcal{H}}{N}\tau^{2}+C_{1}\tau+C_{2}.
\end{equation}
 The constants $C_{1}$and $C_{2}$ are then given by 
 \begin{equation}
C_{1}=\left<\frac{dI}{d\tau}\right>\vert_{\tau=0}=\frac{dI\left(0\right)}{d\tau},\, C_{2}=\left<I\right>\vert_{\tau=0}=I\left(0\right).\label{eq:constants}
\end{equation}
 Thus, according to Ref. \cite{collapse-criteria}, the conditions for the collapse
to occur in a finite time are $\left<d^{2}I/d\tau^{2}\right>\geq0,\, C_{1}\left(0\right)<0,\, C_{2}\left(0\right)>0$
and that the angular momentum integral vanishes, $J\left(\tau=0\right)=0.$
Furthermore, if $C_{1}=0,$ i.e. $dI/d\tau=8\mathcal{H}\tau$,
then if $\mathcal{H}>0$ an initial waveform stretched along $\xi$-axis
will evolve into a structure stretched along the $\eta$-axis for
sufficiently large times. On the other hand, for $\mathcal{H}<0$,
the wave will be stretched along the $\xi$-axis. The  {detailed} analysis is, however, beyond the scope  of the present 
study, instead we will focus on the numerical evolution of the
nonlocal NLSEs (\ref{eq:g1}) and (\ref{eq:g2}). We find that whatever
may be the initial waveform (e.g., 2D Gaussian or close to the exact
solution), the wave amplitude either decays due to dispersion or blows-up
in a finite time due to nonlinearity, when the wave action to the
initial condition is below or above the threshold. Furthermore,
for a particular value of the power of the initial condition close
to the exact solution, we find that the coherent structures, such
as lumps, can also propagate almost without any change. 
\begin{figure}[btp]
\includegraphics[width=6in,height=4in,trim=0.0in 3.5in 0in 4in]{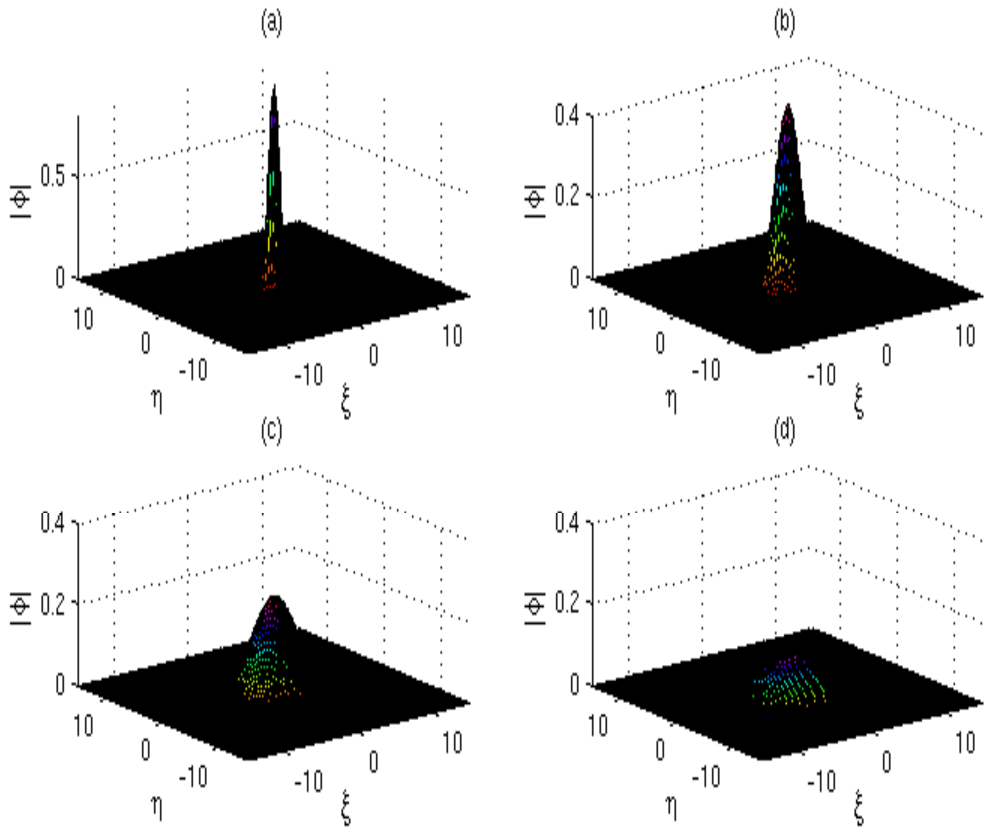}
\caption{(Color online) An initial Gaussian waveform decays in a finite time due to enhancement of the dispersion by
 the static field: (a) $\tau=0$, (b) $\tau=0.5$, (c) $\tau=1$ and (d) $\tau=2$.}

\end{figure}
\begin{figure}[btp]
\includegraphics[width=6in,height=4in,trim=0.0in 3.5in 0in 3.5in]{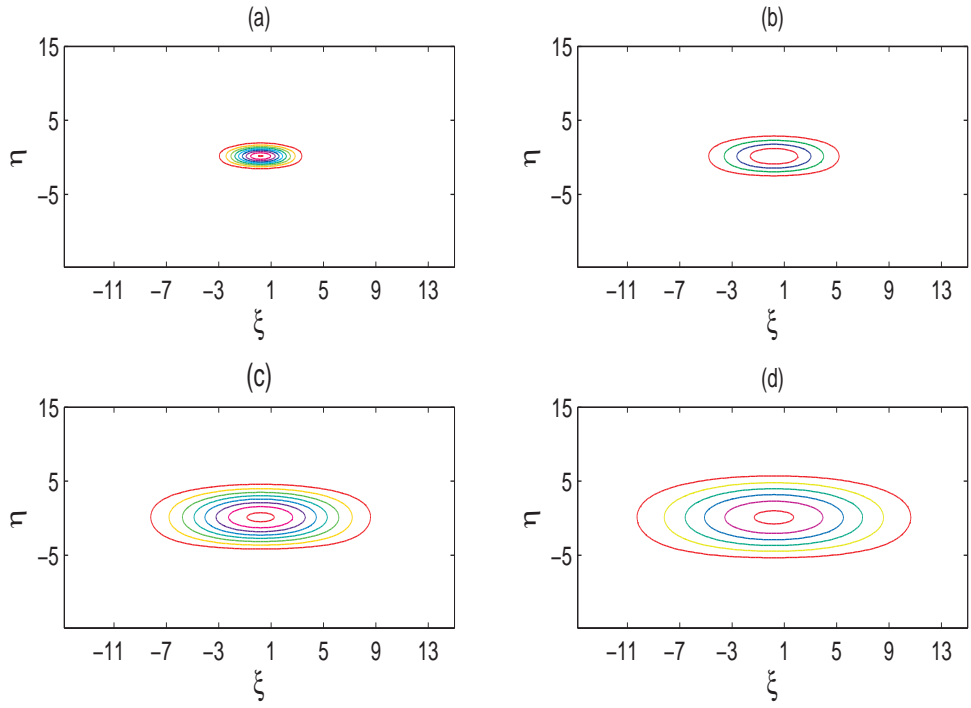}
\caption{(Color online) Contour plots  of the  profiles ($|\phi|=$const.) as in Fig. 3. The  evolution is shown at (a) $\tau=0.5$, (b) $\tau=1$, (c) $\tau=1.5$ and (d) $\tau=2$. }
\end{figure}

The behaviors of the solutions of the DS II equations are quite well-known
and have been investigated by many authors in the context of water
wave propagation with finite depth (see, e.g. Refs. \cite{DS-II-numerical1,DS-II-numerical2}). However, the
solution is still not well understood or almost unknown in the context
of plasma physics. To our knowledge, few authors have  investigated the MI (see, e.g. Refs. \cite{DS-plasma-cylindrical,DS-plasma-MI,DS-II-plasma-dromion}) of a plane wave packet  as well as 
 some analytic solution (see, e.g. \cite{DS-II-plasma-dromion}) of the DS II-like equations in plasmas. 
 
 In our numerical scheme we consider  a  space domain as $\left[-15,15\right]\times\left[-15\times15\right]$
with $150$ grid points in every direction and time step $\delta t=10^{-3}$.
The  parameter values which satisfy the MI condition (\ref{eq:cond1}) are taken as $P_{1}=-0.317$, $P_{2}=0.9327$,  $Q_{1}=-0.1686$, $Q_{2}=0.2876$, $R=0.3028$ and $S=-4.2086$
corresponding to  $k=0.4$ and $n_{0}=2\times10^{33}$ cm$^{-3}$, i.e. $\beta=0.7985$. For
the sake of simplicity, we consider a symmetric Gaussian profile as the initial
condition, i.e. $\phi=A\exp\left(-\xi^{2}/\arrowvert P_{1}\arrowvert-\eta^{2}/P_{2}\right)$. We have tested our numerical results with other forms of initial conditions, namely  $\phi=A/\left(1+\xi^2/\mid P_1\mid+\eta^2/P_2\right)\exp\left(i\xi^{2}/ P_{1}+i\eta^{2}/P_{2}\right)$, which is a localized lump with algebraic decay and also  $\phi=\left[A/\left(1+\xi^{2}/|P_{1}|+\eta^{2}/P_{2}\right)\right]\exp\left(2i\eta/\sqrt{P_{2}}\right)$,  which approximates an exact solution of the DS II-like equations \cite{DS-II-numerical1,DS-II-numerical2},  but observe the similar qualitative features as presented here. 

Figure 3 shows that for $A=1$, the initial wave forms decay with
time, and  go to zero after a  finite time. The corresponding contour plots
are shown in Fig. 4, but for different times. In this case, the presence of the nonlocal static field enhances the dispersion of the profile and  since the quadratic nonlinearities (nonlocal term) are known to be collapse
free, the initial wave action $N$ is not up to the mark for which the cubic nonlinearity
dominates. However, since the effect of the static field $\psi$ is
to modify the nonlinearity originating from the zeroth harmonic modes (or slow modes),
we expect that the coupling of the  field $\psi$ and the wave field amplitude
$\phi$ can drastically affect the evolution dynamics for blowup.
To elucidate it we have considered $A=8$, which is above the critical
power $N$ of blowup for a Gaussian pulse  whose evolution is described
by the 2D NLSE. Thus, Figs. 5 and 6 (contour plot) show that the initial waveform is not preserved, and wave singularity is formed at two points within a finite interval of time leading to double focusing effect. The wave packet thus blows-up  in shorter scales with higher amplitudes. In this case, the self-focusing effects dominate over the dispersion as required. Evolution of such  wave collapse  could be an effective mechanism for energy localization in URD dense plasmas. 
\begin{figure}[btp]
\includegraphics[width=6in,height=4in,trim=0.0in 3.5in 0in 4in]{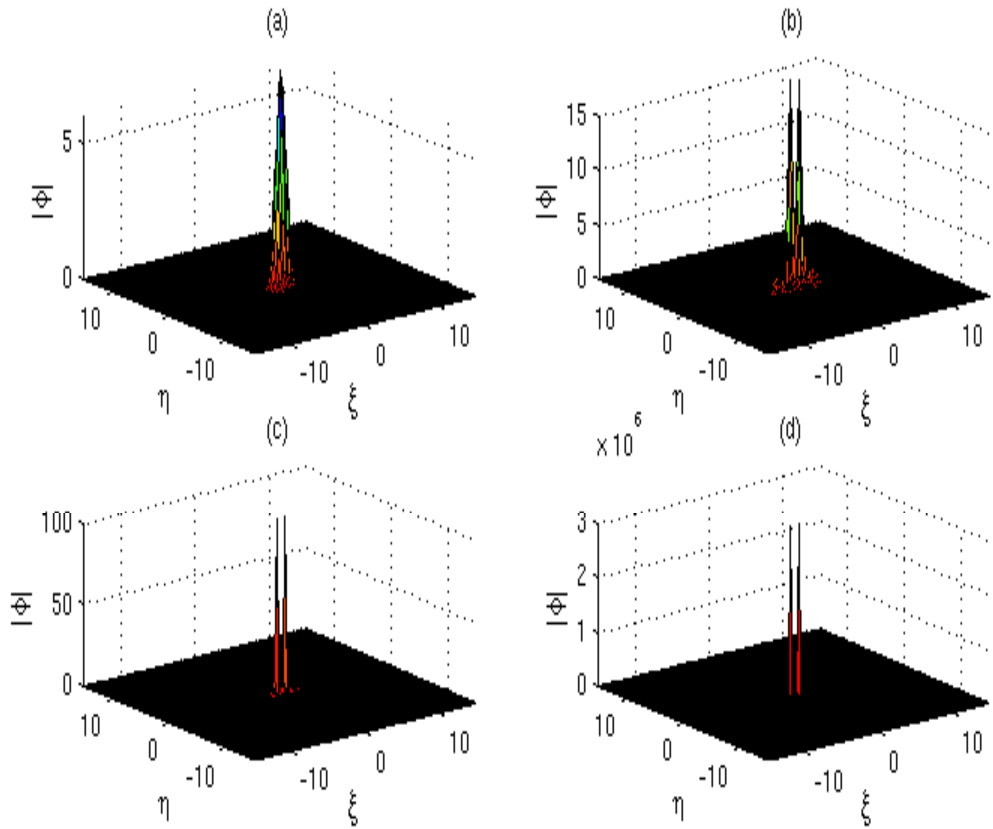} 
\caption{(Color online) Double focusing solutions that blowup in a finite time. The initial wave form is Gaussian with amplitude $A=8$. (a) Initial state: $\tau=0$,
(b) Intermediate state: $\tau=0.35$, (c) Before blowup: $\tau=0.356$ and (d) Shortly after blowup $(|\phi|\sim10^6)$: $\tau=0.358$.} 
\end{figure}
\begin{figure}[btp]
\includegraphics[width=7in,height=3in,trim=0.0in 3.5in 0in 4in]{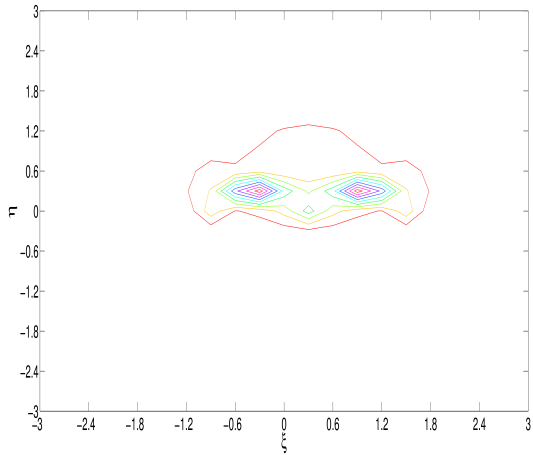} 
\caption{(Color online) Contour plot ($|\phi|=$const.) of a double focusing solution that blows up (as in Fig. 5) is shown at an intermediate time  $\tau=0.35$.}
\end{figure}

A slightly different
perspective is observed when considering the initial profile as \\$\phi=\left[A/\left(1+\xi^{2}/\mid P_{1}\mid+\eta^{2}/P_{2}\right)\right]\exp\left(2i\eta/\sqrt{P_{2}}\right)$, i.e. 
close to the exact solution of DS II equations \cite{DS-II-numerical1,DS-II-numerical2}. Figure 7 shows
that for $A=1$ the coherent structure propagates along the $\xi$-axis almost without any change (after a certain interval)
of its amplitude. We can, however, precisely  {deal with} such  cases by choosing the initial condition as close  to the exact solution of the NLSEs (\ref{eq:g1}) and (\ref{eq:g2}) as well as the proper wave action.
\begin{figure}[btp]
\includegraphics[width=6in,height=4.0in,trim=0.0in 3.5in 0in 4in]{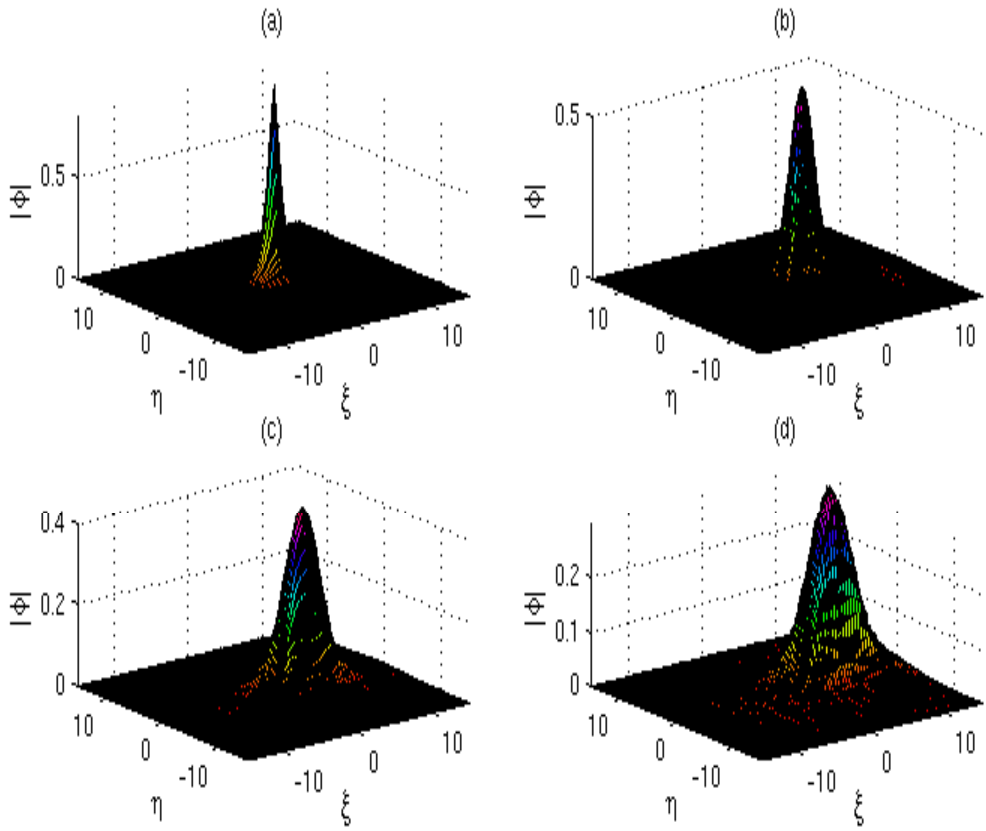} 
\caption{(Color online) Propagating dromions (when the initial waveform is close to the exact solution of the DS-II equations) at different times: (a) $\tau=0$, (b)  $\tau=0.5$, (c)  $\tau=1$ and (d)  $\tau=1.5$.}
\end{figure}
\section{Conclusion}
 We have investigated the multi-dimensional modulation of an electrostatic wave packet propagating in an ultra-relativistic ultra-cold degenerate dense  plasma. The dynamics of such wave packets is described by a coupled set of nonlocal nonlinear Schr{\"o}dinger-like equations that involve a nonlocal (quadratic) nonlinear term. The latter appears due to the static wave field originating from the mean motion (zeroth harmonic) in the plasma. The equations are then used to obtain the instability condition for the modulation of a plane wave packet. It is shown that the density dependent parameter $\beta$, which arises due to the ultra-relativistic pressure of degenerate electrons, shifts the stable (unstable) region at lower density ($n_0\sim10^{30}$ cm$^{-3}$) to unstable (stable) regions at comparatively higher densities ($n_0\sim10^{33}$ cm$^{-3}$), and also that the stable and unstable regions are completely separated in such regimes. The latter can be achievable, e.g. in the interior of massive white dwarfs and neutron stars \cite{Quantum-plasma-review}.  Furthermore, the instability growth rate is obtained and found to be lowered at higher number densities with cut-offs at lower wave numbers of modulation. This implies that the role of $\beta$ is analogous with that due to quantum dispersion associated with the Bohm de Broglie potential for the MI of wave envelopes in quantum plasmas \cite{Misra1,Misra2,Misra3,Misra4}. 
 
We have also shown that the presence of the static field can drastically change the dynamical evolution of the wave packets quite distinctive from the one-dimensional wave packets or 2D NLSE. We  found that when the initial condition is either a Gaussian pulse or a localized lump with algebraic decay,  the  wave amplitudes either disperse away to zero or blowup to infinity at singular points when the wave action (power) is below or above a threshold value. Such a blowup mechanism  could be important for the energy localization in dense plasmas where the electrons are ultra-relativistically degenerate.  On the contrary, when an initial waveform is close to the exact solution of the DS II-like equations \cite{DS-II-numerical1,DS-II-numerical2}, the coherent structure propagates without almost any change. However, confirmation of this behavior needs further detail  {numerical investigation.}  
\acknowledgments{This work was  supported by the Kempe Foundations, Sweden. APM wishes to thank Gert Brodin and Mattias Marklund of Department of Physics, Ume\aa\ University, SE-901 87 Ume\aa ,
Sweden, for their kind help and support.}


\end{document}